\documentclass[conference]{IEEEtran}
\IEEEoverridecommandlockouts
\usepackage{cite}
\usepackage{enumitem} 
\usepackage{comment}
\usepackage{bm} 
\usepackage{amsthm}
\usepackage{multirow}
\usepackage{booktabs}
\usepackage[caption=false,font=footnotesize]{subfig}
\usepackage{dirtytalk}

\newtheorem{remark}{Remark}

\usepackage{amsmath,amssymb,amsfonts}
\usepackage{tikz}
\usepackage{pgfplots}
\usepackage{enumitem}
\pgfplotsset{compat=newest} 
\usepackage{float}   
\usepackage{algorithm}      
\usepackage{algpseudocode}  
\usepackage{graphicx}
\usepackage{textcomp}
\usepackage{xcolor}
\def\BibTeX{{\rm B\kern-.05em{\sc i\kern-.025em b}\kern-.08em
    T\kern-.1667em\lower.7ex\hbox{E}\kern-.125emX}}
\begin{document}
\bstctlcite{IEEEexample:BSTcontrol}

\title{Learning-Based Sensor Scheduling for Delay-Aware and Stable Remote State Estimation}

\author{Nho-Duc Tran, Aamir Mahmood, Mikael Gidlund  \\
\IEEEauthorblockA{Department of Computer and Electrical Engineering, Mid Sweden University, 851 70 Sundsvall, Sweden} Email: \{nhoduc.tran, aamir.mahmood, mikael.gidlund\}@miun.se
\vspace{-10pt}
}


\maketitle


\begin{abstract}

Unpredictable sensor-to-estimator delays fundamentally distort what matters for wireless remote state estimation: not just freshness, but how delay interacts with sensor informativeness and energy efficiency. In this paper, we present a unified, delay‑aware framework that models this coupling explicitly and quantifies a delay‑dependent information gain, motivating an information‑per‑joule scheduling objective beyond age of information proxies (AoI). To this end, we first introduce an efficient posterior‑fusion update that incorporates delayed measurements without state augmentation, providing a consistent approximation to optimal delayed Kalman updates, and then derive tractable stability conditions ensuring that bounded estimation error is achievable under stochastic, delayed scheduling. This conditions highlight the need for unstable modes to be observable across sensors. Building on this foundation, we cast scheduling as a Markov decision process and develop a proximal policy optimization (PPO) scheduler that learns directly from interaction, requires no prior delay model, and explicitly trades off estimation accuracy, freshness, sensor heterogeneity, and transmission energy through normalized rewards. In simulations with heterogeneous sensors, realistic link‑energy models, and random delays, the proposed method learns stably and consistently achieves lower estimation error at comparable energy than random scheduling and strong RL baselines (DQN, A2C), while remaining robust to variations in measurement availability and process/measurement noise. 
\end{abstract}


\section{Introduction}

Wireless remote state estimation is widely used in a broad range of wireless sensor network applications, including industrial monitoring, autonomous systems, and smart infrastructure~\cite{application}. In these settings, numerous heterogeneous sensors monitor and report the system state over bandwidth‑ and energy‑limited links. With such constraints, not every sensor can transmit in every slot; sensor scheduling (deciding who transmits when) is therefore central to achieving accurate, timely estimates under resource constraints. A practical complication is that sensor-to-estimator delays are random and sometimes large, which changes what \textit{good scheduling} entails. Focusing on information freshness alone is insufficient \cite{ao-insufficient}. Performance depends on how freshness, a time property describing how old a packet is when used, interacts with sensor informativeness, a measurement property that explains how much that packet can reduce uncertainty given the sensor’s characteristics and the estimator’s current uncertainty, and with energy cost.

Existing studies on sensor scheduling for remote state estimation typically optimize a single performance axis under simplified assumptions. For instance, accuracy-driven works focus on estimation error alone, often under idealized delay models or single-sensor settings \cite{sensor-schedule-delay-2}. Energy-efficient scheduling has likewise been explored \cite{energy-alone}, but without accounting for how delays and staleness degrade the value of a received packet. Age-of-Information (AoI) formulations capture data freshness \cite{aoi1}, yet they generally overlook random network delays that can render packets outdated upon arrival and do not account for sensor heterogeneity or energy cost. 

While AoI equals the packet delay at the instant of reception, it is only a freshness proxy. The actual estimation gain depends jointly on the delay $\delta$, the current error covariance $\mathbf{P}$, the sensor’s informativeness and energy usage $E$. In this respect, a plausible option is to schedule by maximizing an information-per-joule criterion $\Delta(\mathbf{P},\delta)/E$ instead of minimizing AoI. However, to the best of our knowledge, despite significant progress in related domains, a comprehensive scheduling framework that jointly optimizes estimation accuracy, freshness, heterogeneity, energy use, and stochastic delays has not yet been established



To address these gaps, in this work, we study a multi-sensor remote estimation problem in which heterogeneous sensors, differing in measurement model and noise level, observe a linear dynamical process over wireless links with stochastic sensor-to-estimator delays. We formulate a scheduling problem that explicitly accounts for estimation error, per-transmission energy, and the age-induced attenuation of measurement value via a delay-dependent information gain normalized by energy. Because scheduling in this setting is inherently combinatorial and dynamic, traditional optimization approaches become intractable.. We therefore adopt a reinforcement-learning (RL) formulation in which a policy selects, at each slot, which sensor (if any) should transmit to optimize a long-term objective that jointly balances estimation accuracy, energy consumption, and freshness. The main contributions of this paper are as follows:

\begin{enumerate}
    \item We develop a delay-aware system model that captures the trade-offs among estimation accuracy, packet freshness, sensor heterogeneity, and energy consumption. 
    \item We derive stability conditions that yield the feasible bounded estimation error under delayed and stochastic scheduling, providing theoretical assurance of estimator reliability. 
    \item We design a reinforcement learning scheduler, based on PPO, where the state encodes estimation uncertainty and delay history, and the reward explicitly balances estimation improvement and transmission energy.
    \item Through extensive simulations, we show that the proposed method outperforms random scheduling and other RL baselines such as DQN and A2C, with PPO achieving higher stability and performance under heavy delays. 
\end{enumerate}

The rest of the paper is organized as follows. Section~\ref{section-II} describes the system model and problem formulation. Section~\ref{section-III} details our approach to handling delayed measurements. The feasible stability condition and proposed scheduling algorithm are presented in Section~\ref{section-IV}, and Section~\ref{section-V} provides a thorough discussion of the simulation results. Finally, Section~\ref{section-VI} concludes the paper.
\vspace{-8pt}
\section{System Model}
\label{section-II}

\begin{figure}
\centering
\subfloat[Each sensor $i \in \mathcal{M}$ has a measurement matrix $\mathbf{C}_i$. 
With probability $p_i$, it generates a noisy measurement with covariance $\mathbf{R}_i$.  The most recent measurement, possibly delayed by $\delta_k^i$, is sent to the estimator at energy cost $E^i$ if the invitation is received at time step $k$. 
]{
      \resizebox{0.7\linewidth}{!}{
        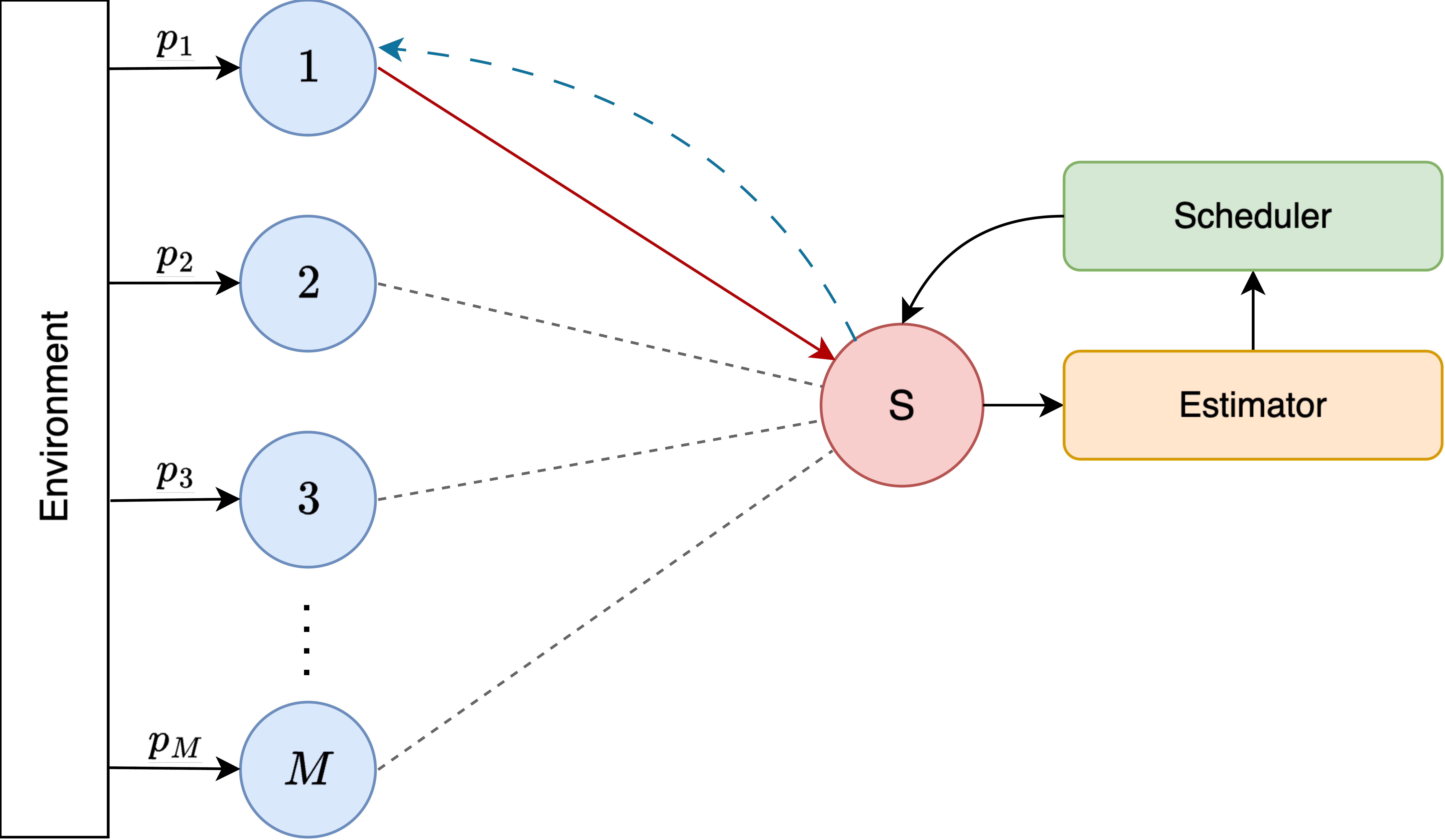
      }
    }\\
    \vspace{-7pt}
    \subfloat[Due to the randomness of new sample generation, the delay may vary depending on when sensor $i$ is selected.]{
    \resizebox{0.9\linewidth}{!}{
        \begin{tikzpicture}[font=\small,>=stealth,thick,x=1.0cm,y=1.0cm, line width=1pt]

  \draw (0,4.5) -- (7,4.5) node[right]{Estimator};
  \draw (0,2.25) -- (7,2.25) node[right]{Sensor $i$};

  \foreach \x in {0,1.4,2.8,4.2,5.6,7} {
    \draw[dashed,gray!70] (\x,2.25) -- (\x,4.5);
  }

  \draw[->,blue, line width=1pt] (0,1.8) -- (0,2.25);     
  \draw[->,blue, line width=1pt] (2.8,1.8) -- (2.8,2.25); 
  \draw[->,blue, line width=1pt] (4.2,1.8) -- (4.2,2.25); 

  \draw[->,cyan,dashed, line width=1pt] (1.4,4.5) -- (1.7,2.25); 
  \draw[->,cyan,dashed, line width=1pt] (4.2,4.5) -- (4.5,2.25); 
  \draw[->,cyan,dashed, line width=1pt] (5.6,4.5) -- (5.9,2.25); 

  \draw[->,red, line width=1pt] (1.7,2.25) -- (2.0,4.5);  
  \draw[->,red, line width=1pt] (4.5,2.25) -- (4.8,4.5);  
  \draw[->,red, line width=1pt] (5.9,2.25) -- (6.2,4.5);  

  \draw[<->,magenta,thick, line width=1pt] (0,1.6) -- (2.0,1.6);   
  \draw[<->,magenta,thick, line width=1pt] (4.2,1.3) -- (4.8,1.3); 
  \draw[<->,magenta,thick, line width=1pt] (4.2,1.6) -- (6.2,1.6); 

  \node[
    draw=white,
    fill=white,
    rounded corners,
    anchor=west,
    xshift=1.2cm,
    yshift=-1.2cm,
    inner sep=4pt
  ]
  at (7.0,4.5)
  {
    \begin{tabular}{@{}l@{\ }l}
      \tikz[baseline=-0.8ex]{\draw[->,cyan,dashed, line width=1pt] (0,0) -- (0.5,0);} & \;Invitation\\
      \tikz[baseline=-0.8ex]{\draw[->,red, line width=1pt] (0,0) -- (0.5,0);} & \;Sending measurement\\
      \tikz[baseline=-0.8ex]{\draw[->,blue, line width=1pt] (0,0) -- (0.5,0);} &\;New sample\\
      \tikz[baseline=-0.8ex]{\draw[<->,magenta,thick, line width=1pt] (0,0) -- (0.5,0);} & \;Delay
    \end{tabular}
  };

\end{tikzpicture}
        }
    }
\caption{Overview of network model and optimization problem. The question is which sensor (or none) is the best action at each time step.}
\vspace{-15pt}
\label{fig:system_model}
\end{figure}
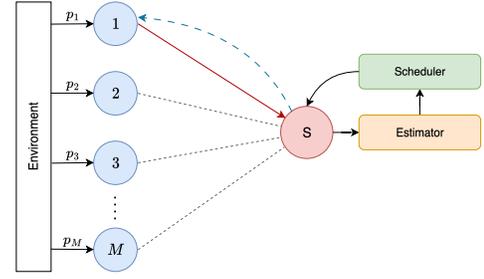
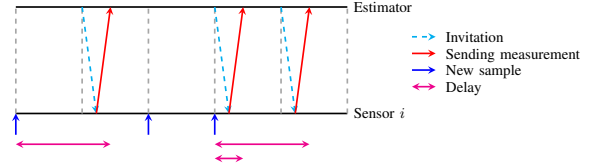

As shown in Fig.~\ref{fig:system_model}, we consider a time-slotted wireless sensor network where $M$ sensors monitor $N$ states. Let $\mathcal{M}=\{1,\dots,M\}$ and $\mathcal{N}=\{1,\dots,N\}$ denote the sensor and state indices, respectively. The system state is $\mathbf{x}_k = [x_1(k),\dots,x_N(k)]^\top \in \mathbb{R}^N.$ Each sensor $i \in \mathcal{M}$ observes $m_i$  features of $\mathbf{x}_k$ ($1 \leq m_i\leq N$),  and generates a measurement $\mathbf{y}_k^i \in \mathbb{R}^{m_i}$ with probability $p_i$ every time step. Since a single channel is shared, at most one sensor transmits per slot, determined by the scheduling signal from the remote estimator. When selected, sensor $i$ sends its latest measurement, possibly acquired at $k-\delta$ ($\delta \ge 0$), and such delays can degrade estimation accuracy.

\subsection{State Space Model}
We adopt a discrete linear time-invariant (LTI) process–measurement model as the canonical abstraction for sampled-data systems. It provides the standard, analytically tractable baseline for rigorously characterizing delayed estimation and scheduling, including closed-form covariance evolution and stability conditions based on the observability of unstable modes. The posterior-fusion mechanism and the policy-learning approach extend to nonlinear processes via Extended Kalman Filter/Unscented Kalman Filter linearization, but we state formal guarantees for the LTI case, consistent with common practice in sampled-data control and estimation \cite{LTI-2}. 
For an $N$-dimensional LTI system observed by $M$ heterogeneous sensors, the state $\mathbf{x}_k$ evolves as
\begin{align}
  \mathbf{x}_{k+1} &= \mathbf{A}\,\mathbf{x}_k + \mathbf{w}_k,\quad \mathbf{w}_k \sim \mathcal{N}(0,\,\mathbf{Q}), \label{eq:state}\\[3pt]
  \mathbf{y}_k^i &= \mathbf{C}_i\,\mathbf{x}_k + \mathbf{v}_k^i,\quad \mathbf{v}_k^i \sim \mathcal{N}(0,\,\mathbf{R}_i), \label{eq:measurement}
\end{align}
where \(\mathbf{A}\in\mathbb{R}^{N\times N}\) is the state transition matrix, \(\mathbf{C}_i\in\mathbb{R}^{m_i\times N}\) is the measurement matrix of sensor \(i\), while
the process noise \(\mathbf{w}_k\) and measurement noise \(\mathbf{v}_k^i\) are uncorrelated zero-mean Gaussian vectors with covariances
$
\mathbb{E}[\,\mathbf{w}_k \mathbf{w}_j^\top] = \delta_{kj} \,\mathbf{Q},\quad
\mathbb{E}[\,\mathbf{v}_k^i (\mathbf{v}_j^i)^\top] = \delta_{kj}\,\mathbf{R}_i,
$
where \(\delta_{kj} = 1\) if \(k=j\) and \(0\) otherwise. 



\subsection{Scheduler}

We define the binary variable \(\gamma_k^i\) to indicate whether the measurement \(\mathbf{y}_k^i\) from sensor \(i\) at time \(k\) is chosen:
\begin{equation}
\label{eq:decision_var}
\gamma_k^i = 
\begin{cases}
1, & \text{if } \mathbf{y}_k^i \text{ is selected},\\
0, & \text{if } \mathbf{y}_k^i \text{ is not selected}.
\end{cases}
\end{equation}

At each time step $k$, the selection of sensors is captured by the vector
$
\boldsymbol{\gamma}_k \;=\; \bigl[\gamma_k^1 \;\;\gamma_k^2\;\;\dots\;\;\gamma_k^M\bigr].
$ Therefore, the general decision can be represented as
$
\theta = \{ \boldsymbol{\gamma}_k\}_{k=1}^{\infty}.
$
\vspace{-8pt}
\subsection{Estimation Model and Error Covariance}
Denote the prior estimate of \(\textbf{x}_{k}\) as \(\hat{\textbf{x}}_{k|k-1}\), and \(\hat{\textbf{x}}_{k|k}\) as the posterior estimate of \(\textbf{x}_{k}\) after updating the measurement \(\textbf{y}_{k}\): 
\setlength{\abovedisplayskip}{4pt}
\begin{align}
\hat{\textbf{x}}_{k|k-1} \triangleq \mathbb{E}[\textbf{x}_{k} | \textbf{y}_{k-1}], \;\;
\hat{\textbf{x}}_{k|k} \triangleq \mathbb{E}[\textbf{x}_{k} | \textbf{y}_{k}]. 
\end{align}

In practical scenarios, acquiring the true system state for evaluating performance and making decisions is often infeasible. Therefore, we rely on the estimation error covariance matrix $\mathbf{P}$ as a proxy for estimation accuracy. Specifically,  $\mathbf{P}$ with respect to $\hat{\mathbf{x}}_{k|k-1}$ and $\hat{\mathbf{x}}_{k|k}$ is given as
\begin{align}
    \mathbf{P}_{k|k-1} &= \mathbb{E}\bigl[ (\mathbf{x}_k - \hat{\mathbf{x}}_{k|k-1})(\mathbf{x}_k - \hat{\mathbf{x}}_{k|k-1})^\top\bigr], \label{cov1}\\
    \mathbf{P}_{k|k} &= \mathbb{E}\bigl[ (\mathbf{x}_k - \hat{\mathbf{x}}_{k|k})(\mathbf{x}_k - \hat{\mathbf{x}}_{k|k})^\top\bigr].
    \label{cov2}
\end{align}

\subsection{Energy Consumption Model}

In remote state estimation, heterogeneous and dispersed devices incur different costs for sending data to the remote estimator.  We define the transmission-cost vector $\mathbf{E} = [E^{1}, E^{2}, \dots, E^{M}],$ where $E^i$ follows the energy-limited sensor model. 
The total energy for transmitting $N_b$ bits is~\cite{energy1}
\begin{equation}
\label{eq:Etot}
E^i = (P_t + P_c)\,
\frac{N_b}{B \log_{2}\!\bigl(1+\rho_i\bigr)}, 
\qquad
\rho_i = \frac{\eta P_t G_i}{N_{0}B},
\end{equation}
with $P_t$: transmit power, $P_c$: circuit/baseband power, $B$: bandwidth, $G_i$: channel gain, $N_0$: noise spectral density, $\rho_i$: SNR of sensor $i$, and $\eta$: PA efficiency.

\subsection{Problem Formulation}
We aim to jointly optimize state estimation accuracy and sensor energy consumption in the presence of uncertain measurement delays by designing a scheduling policy~$\theta$ that minimizes the long-term average cost. Because these two metrics differ in scale and units, a direct combination would bias the optimization. To eliminate this imbalance, we apply min–max normalization to map both estimation error and energy cost into the range $[0,1]$, leading to the formulation:
\begin{align}
\min_{\theta} \;\; \lim_{T \rightarrow \infty} \frac{1}{T}\sum_{k=1}^{T} \left\{ 
\dfrac{\mathrm{trace}(\mathbf{P}_k)}{\mathrm{trace}(\mathbf{P}_0)} 
+ \beta\, \bm{\gamma}_k\, \frac{E_k}{\max{\mathbf{E}}}^\top \right\}.
\label{eq:problem_normalize}
\end{align}
In this study, the weighting parameter $\beta$ is assumed to be fixed, while a general framework for multi-objective balancing is left for future work. Importantly, however, solving~\eqref{eq:problem_normalize} requires evaluating how delayed measurements influence the evolution of $\mathbf{P}_k$. Accordingly, the next section develops a delay-aware posterior-fusion estimator and a delay-dependent information gain that will later drive scheduling decisions.

\section{Delay-Aware State Estimation Methods}
\label{section-III}

We begin with the standard Kalman prediction and update, then develop a delay-aware posterior-fusion estimator that incorporates delayed measurements without replay.

\subsection{State Estimation}

Every timestep, the estimator predicts the next state and error covariance of the system by
\begin{align}
  \hat{\mathbf{x}}_{k|k-1} &= \mathbf{A}\,\hat{\mathbf{x}}_{k-1|k-1}, \label{eq:pred_state}\\[1mm]
  \mathbf{P}_{k|k-1} &= \mathbf{A}\,\mathbf{P}_{k-1|k-1}\,\mathbf{A}^\top + \mathbf{Q}. \label{eq:pred_cov}
\end{align}

When a timely measurement \(\mathbf{y}_k^i\) is available at time \(k\), the filter calculates the Kalman gain, the posterior estimate state, and error covariance
\begin{align}
  \mathbf{K}_k &= \mathbf{P}_{k|k-1}\,\mathbf{C}_i^\top \bigl(\mathbf{C}_i\,\mathbf{P}_{k|k-1}\,\mathbf{C}_i^\top + \mathbf{R}_i\bigr)^{-1}, \label{eq:kalman_gain}\\
  \hat{\mathbf{x}}_{k|k} &= \hat{\mathbf{x}}_{k|k-1} + \mathbf{K}_k\bigl(\mathbf{y}_k^i - \mathbf{C}_i\,\hat{\mathbf{x}}_{k|k-1}\bigr), \label{eq:state_update}\\[1mm]
  \mathbf{P}_{k|k} &= \bigl(\mathbf{I} - \mathbf{K}_k\,\mathbf{C}_i\bigr) \mathbf{P}_{k|k-1}. \label{eq:cov_update}
\end{align}

\subsection{Incorporating Delayed Measurement Updates}

At time $k$, suppose a measurement from sensor $i$ generated at 
$\tau = k-\delta_{i,k}$ arrives is:
$
y^i_{\tau} = \mathbf{C}_i \mathbf{x}_{\tau} + \mathbf{v}^i_{\tau}, 
\quad 
\mathbf{v}^i_{\tau}\sim \mathcal{N}(0,\mathbf{R}_i).
$


\emph{1) Update at the generation time.}  
From the stored prior $(\hat{\mathbf{x}}_{\tau},\mathbf{P}_{\tau})$, 
we apply a standard Kalman update in \eqref{eq:kalman_gain}-\eqref{eq:cov_update} to find posterior $(\hat{\mathbf{x}}_{\tau|\tau}^{(i)},\mathbf{P}_{\tau|\tau}^{(i)})$.

\emph{2) Step-by-step propagation and fusion.}  
At each ensuing time instant $j$ ($\tau< j < k$), the posterior $(\hat{\mathbf{x}}_{\tau|\tau}^{(i)},\mathbf{P}_{\tau|\tau}^{(i)})$ is advanced forward by employing the prediction mechanisms \eqref{eq:pred_state}-\eqref{eq:pred_cov}. If another sensor's measurement has already been integrated at $j$, the advanced state $(\hat{\mathbf{x}}_{j}^{p},\mathbf{P}_{j}^{p})$ is amalgamated with the system's posterior $(\hat{\mathbf{x}}_{j|j},\mathbf{P}_{j|j})$:
\begin{align}
\mathbf{K}_j &= \mathbf{P}_{j|j}^p \big(\mathbf{P}_{j|j}^p + \mathbf{P}_{j|j}\big)^{-1}, \\
\hat{\mathbf{x}}_{j|j}^r &= \hat{\mathbf{x}}_{j|j}^p + 
\mathbf{K}_j \big(\hat{\mathbf{x}}_{j|j} - \hat{\mathbf{x}}_{j|j}^p\big), \\
\mathbf{P}_{j|j}^r &= (\mathbf{I}-\mathbf{K}_j)\mathbf{P}_{j|j}^p. \label{eq:delay_update}
\end{align}

Instead of executing the filter from $\tau$ again by replaying delayed measurements, which, though optimal, remains impractical, our approach utilizes each posterior $(\hat{\mathbf{x}}_{j|j},\mathbf{P}_{j|j})$ to represent all input data up to $j$ in a Gaussian configuration. Considering this posterior as a virtual observation characterized by mean $\hat{\mathbf{x}}_{j|j}$ and covariance $\mathbf{P}_{j|j}$, we can integrate it with the delayed state, thus replaying data in a compressed form. Consequently, posterior fusion provides a practical and efficient approximation of complete updates, effectively utilizing all prior information without reprocessing.

\emph{3) Final estimate at time $k$.}  
After iterating the propagate--fusion procedure until $k$, the corrected estimate 
$(\hat{\mathbf{x}}_{r|k},\mathbf{P}_{r|k})$ is obtained. Its effect is quantified as
$
\Delta_i(\mathbf{P}_{k},\delta_{i,k}) 
= \mathrm{trace}(\mathbf{P}_{k|k-1}) - \mathrm{trace}(\mathbf{P}_{r|k}),
$
which monotonically decreases with larger $\delta_{i,k}$.

\vspace{-5pt}
\section{Stability Analysis and Learning-Based Scheduling under Uncertain Delays}
\label{section-IV}

\subsection{Stability Analysis}
\label{sec:stability}

We say stabilization of the estimation error is \emph{feasible} if there exists an admissible schedule $\theta$ such that
\[
G = \sup_{\mathbf{P}_0 \geq 0} \limsup_{k \to \infty} \left\| \mathbb{E} \left( \mathbf{P}_k  \right) \right\| < \infty.
\]

Assuming, without loss of generality, that the state transition matrix $\mathbf{A}$ is already in Jordan normal form, $\mathbf{A} = \mathbf{I}^{-1}\mathbf{A}\mathbf{I}=\mathrm{diag}(\mathbf{A}_u, \mathbf{A}_s)$, then if $\rho(\mathbf{A}) < 1$, the system exhibits asymptotic stability even in the absence of measurement updates \cite{dansimon}. Conversely, let us consider $\rho(\mathbf{A}_u)\ge 1$, and denote $ \mathbf{\bar{C}}_i:=\mathbf{C}_i\mathbf{I}=[\,\mathbf{C}_i^u\;\;\mathbf{C}_i^s\,]$. The detectability of the system requires that its unstable modes be observable, which is equivalent to the existence of a schedule $\boldsymbol{\theta} = \{\boldsymbol{\gamma}_1, \boldsymbol{\gamma}_1 , \ldots, \boldsymbol{\gamma}_r \}, \,r = \dim(\mathbf{A_u})$ such that matrix $ \begin{bmatrix}
\mathbf{C}^u_{\boldsymbol{\gamma}_1}, \; \mathbf{C}^u_{\boldsymbol{\gamma}_2}\mathbf{A}_u , \; \ldots , \; \mathbf{C}^u_{\boldsymbol{\gamma}_r}\mathbf{A}_u^{r-1}
\end{bmatrix}^\top \text{ has full rank.}$

\begin{remark}
\label{lemma}
Define $\mathcal{O} =
\begin{bmatrix}
\mathbf{B}_0, \, \mathbf{B}_1, \,\ldots , \,  \mathbf{B}_{r-1}
\end{bmatrix}^\top 
\text{ where }
\mathbf{B}_i =
\begin{bmatrix}
\mathbf{C}^u_1 , \, \mathbf{C}^u_1, \,\ldots , \, \mathbf{C}^u_M
\end{bmatrix}^\top
\mathbf{A}^i_u .
$ The unstable modes $u$ of the system is observable only when

\begin{enumerate}[label=\arabic*), leftmargin=1.2cm]
  \item $\operatorname{rank}(\mathcal{O}) = r$.
  
  \item There exist vectors $\mathbf{b}_i \in \operatorname{row}(\mathbf{B}_k)$ for each 
  $i \in \{0,\dots,r-1\}$ such that $\{\mathbf{b}_0,\mathbf{b}_1,\dots,\mathbf{b}_{r-1}\}$ are linearly independent.  Equivalently, for all $i$,
  \vspace{-5pt}
  \[
  \mathcal{R}_k = \operatorname{row}(\mathbf{B}_i) \not\subseteq \sum_{j\neq i} \mathcal{R}_j,
  \quad
  \dim\!\left(\sum_{i=0}^{r-1} \mathcal{R}_i \right) = r.
  \]
\end{enumerate}
\end{remark}



\subsection{Problem Analysis and Design Rationale}


\subsubsection{Estimation Accuracy Dependencies}
The error covariance reduction when an update is received after a delay $\delta_k^i$ from sensor $i$ is 
$
\Delta_i(\mathbf{P}_{k}) 
= \mathbf{P}_{k|k-1} - \mathbf{P}_{k|k}.
$ Thus, $\Delta_i(\mathbf{P}_{k}, \delta)$ exhibits a non-increasing behavior in relation to $\delta$ and is bounded below by 0 in the absence of any updates.
$
\Delta_i(\mathbf{P}_{k},\delta_1) \ge \Delta_i(\mathbf{P}_{k},\delta_2) > 0, \quad \delta_1 < \delta_2.
$
The $\Delta_i(\mathbf{P}_{k})$ also depends on $\mathbf{C}_i$ and $\mathbf{R}_i$. A sensor with large $\mathbf{C}_i$ rank or small noise $\mathbf{R}_i$ provides higher informativeness. Conversely, redundant or noisy sensors produce small $\Delta_i$. 

\subsubsection{Energy Consumption Dependencies}
The transmission energy of sensor $i$ is fixed as $E^i$ in \eqref{eq:Etot}. The efficiency of choosing sensor $i$ is characterized by the benefit-to-cost ratio
$
\zeta_i(\mathbf{P}_{k}, \delta_k^i) 
= \Delta_i(\mathbf{P}_{k}, \delta_k^i)/\beta E^i.
$
For larger $\delta_k^i$, the numerator $\Delta_i$ shrinks, while different $\mathbf{C}_i,\mathbf{R}_i$ yield different $\Delta_i$, and $E^i$ also varies across sensors.  


\subsubsection{Accuracy–Energy Trade-off}  
Optimizing \eqref{eq:problem_normalize} therefore requires balancing:
$
\max_i \;\zeta_i(\mathbf{P}_{k}, \delta_k^i) 
= \max_i \;\Delta_i(\mathbf{P}_{k}, \delta_k^i)/\beta E^i.
$
This shows mathematically that freshness, heterogeneity, and random delays enter the problem implicitly, and that the optimal policy must jointly balance accuracy and energy through their combined effect.

\vspace{-5pt}

\subsection{MDP Formulation of the Scheduling Problem}

We formulate the sensor scheduling problem as a Markov decision process (MDP) characterized as:
\begin{itemize}[leftmargin=*]
    \item \textit{State Space}: 
    At time step $k$, the state includes (i) the logarithm of the diagonal entries of the error covariance matrix, and (ii) the most recent $\nu$ sensor–delay pairs received by the estimator. Formally, $s_k = \big(\log\!\big(\mathrm{diag}(\mathbf{P}_{k-1}) + \epsilon\big), \; (i,\delta)_{k-1:k-\nu}\big),$
    where $(i,\delta)_{k-1:k-\nu}$ denotes the sequence of the last $\nu$ sensor identifiers $i$ and their associated delays $\delta$. 
    The state space is thus
    $
    \mathcal{S} = (\mathbb{R}^+)^N \times (\{0,1,\dots,M\} \times \mathbb{R}^+)^\nu .
    $


    \item \textit{Action Space}: 
    At each time step $k$, the scheduler selects exactly one sensor or remains idle. 
    We model this as a categorical action $\mathcal{A} = \{0,1,\dots,M\}$, where $a_k=0$ denotes idle and $a_k=i$ denotes scheduling sensor $i$. 

    \item \textit{Reward}:
    The reward at time step $k$ is defined as $r_k = - \hat{u}_k \;-\; \beta \hat{e}_k$,
    where $\hat{u}_k$ denotes the normalized uncertainty, $\hat{e}_k$ the normalized energy consumption at time step $k$: $\hat{u}_k = \mathrm{trace}(\mathbf{P}_k)/\mathrm{trace}(\mathbf{P}_0), \, \hat{e}_k = E_k/\max{\mathbf{E}}.$ This formulation encourages the agent to minimize long-term estimation uncertainty while balancing the trade-off with energy consumption. The reward formulation implicitly enforces the stability feasibility condition: when observability of unstable modes is lost, the error covariance diverges, and the resulting large uncertainty term leads to strongly negative rewards. 
    

    \item \textit{Transition Dynamics}: The probability of state transition, as referenced in $\mathbb{P}(s_{k+1}|s_k, a_k)$, can be derived directly from the state updating rules delineated in \eqref{eq:decision_var}, \eqref{eq:pred_state}-\eqref{eq:delay_update}, in conjunction with the probabilistic assessments of each sensor as presented in $p_i$.
\end{itemize}
The state $s_k$ captures (i) estimation uncertainty via the log of the covariance diagonal, ensuring scale invariance and avoiding dominance by large variances, and (ii) delay-awareness via the last $v$ sensor–delay pairs, summarizing recent staleness and scheduling history in a compact form.


\subsection{Proximal Policy Optimization}
Given the inherently continuous nature and high dimensionality of the state space, we utilize the proximal policy optimization (PPO) algorithm \cite{ppo} to learn a stochastic policy, for its robustness and comparatively low hyperparameter sensitivity.
\vspace{5pt}
\subsubsection{The Actor}
The policy is represented by a neural network $\pi_{\theta}(a_t | s_t)$, which takes the current state $s_t$ as input and outputs a probability distribution over possible actions that reflect equivalent accumulated reward in the future.

\subsubsection{The Critic}
The value network $V_{\phi}(s_t)$ estimates the state value returns and updates by minimizing the mean squared error (MSE) between predicted $V_{\phi}(s_t)$ values and the empirical return $\hat{R}_t$. 

\subsubsection{The update}
Both networks are trained jointly using the PPO objective, which combines three components: $L(\theta, \phi) = L_{\text{clip}}(\theta) + c_v\,L_{\text{value}}(\phi) - \beta_p\,H[\pi_\theta]$, where $L_{\text{clip}}$ represents the clipped surrogate policy loss, $L_{\text{value}}$ is the value function loss, and $H[\pi_\theta]$ (entropy) encourages exploration to avoid early convergence to deterministic policies. Coefficients $c_v$ and $\beta_p$ adjust the value and entropy weighting. 
 


\section{Results and Discussion}
\label{section-V}
\vspace{-3pt}
\subsection{Experimental Setup}
\label{setup}
We initialize the system described in Section~\ref{section-II} assuming a 5-dimensional state (\(N=5\)) observed by \(M=20\) sensors. 
The state–space parameters are initialized as:
\begin{itemize}
    \item $\mathbf{A} \in \mathbb{R}^{5 \times 5}, \;a_{ij} \sim \mathcal{U}[0,1)  $,
    \item $\mathbf{C}_i \in \mathbb{R}^{m \times 5}, \;c_{ij} \sim \mathcal{U}[-1,1), m \in [1,5]  $,
    \item $\mathbf{Q} = \mathbf{q}\mathbf{q}^\top + \epsilon \mathbf{I}, \; \mathbf{q} \in \mathbb{R}^{5 \times 5}, \;q_{ij} \sim \mathcal{U}[0,1)$
    \item $\mathbf{R}_i = \mathbf{r}\mathbf{r}^\top + \epsilon \mathbf{I}, \; \mathbf{r} \in \mathbb{R}^{m \times m}, \;r_{ij} \sim \mathcal{U}[0,1)$
\end{itemize}


The sensors are independently and uniformly deployed in an annulus with inner radius \(d_{\min}=100\,\mathrm{m}\)
and outer radius \(d_{\max}=300\,\mathrm{m}\). 
To focus on analyzing scheduling/aggregation effects, as in \cite{Sun2025AoISchedulingTAC}, the channel gain is abstracted by the Friis free-space model, ${G_{i} \;=\; \frac{\lambda^{2}}{(4\pi\,d_i)^{2}}\;G_{t}\,G_{r}}$,
with $\lambda$ the wavelength, $d_i$ the sensor--receiver separation for sensor~$i$, and $G_{t}$, $G_{r}$ the transmit and receive antenna gains, respectively. Transmit powers are set to meet a target minimum SNR at the edge so that all scheduled packets are reliably received.
Table~\ref{table1} summarizes the wireless link-related simulation parameters.

\subsection{Policy Learning Process}
\subsubsection{Hyperparameter tuning}
We tuned the hyperparameters using the Optuna optimization framework~\cite{optuna_2019}. A total of $50$ trials, each comprising $10,000$ steps, were executed to determine the optimal parameter set that yields the highest accumulated reward over $10$ evaluation-phase episodes. Table \ref{tab:ppo-hyperparams} enumerates the tuned parameters, their respective ranges, and the achieved optimal values.

\begin{table}[tp]
\centering
\caption{Parameters for Simulation Setup \cite{energy1}}
\vspace{-6pt}
\label{tab:SimParams}
\resizebox{0.75\columnwidth}{!}{ 
\begin{tabular}{lcl}
\hline
\textbf{Parameter} & \textbf{Symbol} & \textbf{Value} \\
\hline
Number of bits to transmit & $N_b$ & $280$ \\
Bandwidth                  & $B$ & $2~\mathrm{MHz}$ \\
Transmit/receive-antenna gains   & $G_t, G_r$ & $1$ \\
Wavelength                 & $\lambda$ & $0.125~\mathrm{m}$ \\
Distance                   & $d_i$ & $\mathcal{U}(100, 300)~\mathrm{m}$ \\
Noise spectral density     & $N_{0}$ & $-174~\mathrm{dBm/Hz}$ \\
Minimum SNR                 & $\rho$ & $10~\mathrm{dB}$ \\
Amplifier efficiency       & $\eta$ & $0.8$ \\
Circuit/baseband power     & $P_c$ & $10~\mathrm{mW}$ \\
Transmit power     & $P_t$ & $10~\mathrm{mW}$ \\
Total time step     & $T$ & $100$ \\
Weight coefficient     & $\beta$ & $0.1$ \\
Sample probability     & $p_i$ & $\mathcal{U}(0.4, 0.6)$ \\
\hline
\end{tabular}
}
\label{table1}
\vspace{-12pt}
\end{table}

\begin{table}[t]
\centering
\caption{PPO hyperparameter search space and selected values.}
\vspace{-6pt}
\resizebox{0.75\columnwidth}{!}{
\begin{tabular}{l l c} 
\toprule
\textbf{Hyperparameter} & \textbf{Search Range} & \textbf{Optimal Value} \\
\midrule
Learning rate    & $[10^{-8}, 10^{-3}]$  & $1.9 \times 10^{-4}$ \\
Discount factor & $[0.90, 0.99]$             & $0.94$ \\
Clip coefficient & $[0.01, 0.5]$                   & $0.18$ \\
Entropy coefficient & $[10^{-4}, 2 \times 10^{-2}]$ & $0.01$ \\
Number of environments & $\{4, 8, 16, 32\}$         & $16$ \\
Number of steps & $[64, 2048]$, step $=64$         & $192$ \\
Number of minibatches & $[4, 64]$, step $=4$    & $28$ \\
Update epochs   & $[4, 128]$, step $=8$            & $108$ \\
GAE $\lambda$   & $[0.80, 0.99]$                   & $0.98$ \\
\bottomrule
\end{tabular}
}
\label{tab:ppo-hyperparams}
\vspace{-15pt}
\end{table}

\subsubsection{Training phase}

\begin{figure}[t]
    \centering
    \includegraphics[width=0.9\linewidth]{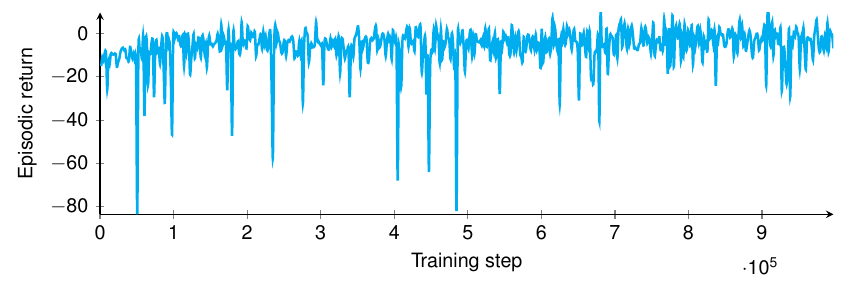}
    \vspace{-10pt}
    \caption{PPO Reward learning curve}
    \label{fig:reward-learning}
    \vspace{-10pt}
\end{figure}
\begin{figure}[tb]
    \centering
\includegraphics[width=1.0\linewidth]{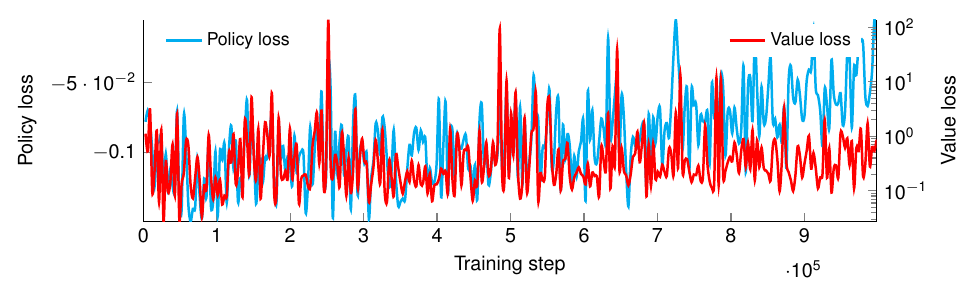}
        \vspace{-20pt}
    \caption{PPO training loss}
    \vspace{-14pt}
    \label{fig:loss-training}
\end{figure}

At each iteration, $N$ parallel environments generate rollouts of length $T$, forming a batch of $N \times T$ samples. The actor–critic networks process these trajectories, each implemented as a two-layer MLP with 128 and 64 hidden units, respectively, to produce action probabilities and value estimates.  Adam is used as the optimizer, and training proceeds for $10^6$ environment steps. As shown in Fig.~\ref{fig:reward-learning}, the reward curve improves steadily with bounded fluctuations, reflecting stable learning. Fig.~\ref{fig:loss-training} shows that the policy loss oscillates around zero while the value loss converges after an initial transient, confirming that PPO achieves reliable and stable convergence in our setting.

\vspace{-2pt}
\subsection{Scheduling Policy Performance}
We evaluate the performance of the proposed PPO-based scheduling policy against three baselines: random, Deep Q-Network (DQN)-based, and Advantage Actor–Critic (A2C)-based. To account for the stochastic nature of the environment and random packet delays, each method is simulated over $10^3$ independent runs with the same random seeds. 

\subsubsection{Baseline Comparison}
As shown in \textit{Standard} column of Table \ref{tab:comparison}, PPO achieves the lowest mean cost and smallest variance of the objective value, indicating both efficiency and stability. PPO naturally learns to maintain bounded estimation error, operating within the stability-feasible region derived in Section IV-A. In contrast, random scheduling performs worst, as expected for its delay-energy-agnostic scheduling approach. Interestingly, while DQN approaches PPO in mean performance, its variance remains much higher, confirming its instability under stochastic delays. A2C provides more consistent results than DQN but falls short of PPO in both accuracy and robustness. See Sec.~\ref{sec:paramterVariations} and the remaining Table III columns for parameter-variation results.


\begin{table*}[tb]
\centering
\caption{Performance comparison of scheduling methods under parameter variations. Each element represents the mean and standard deviation over 1000 simulation runs.}
\vspace{-8pt}
\label{tab:comparison}
\resizebox{0.9\textwidth}{!}{%
\begin{tabular}{lcccccccc}
\toprule
\multirow{2}{*}{\textbf{Method}} & \multirow{2}{*}{\textbf{Standard}} & \multicolumn{2}{c}{\textbf{$p_i$ variation}} & \multicolumn{2}{c}{\textbf{$q_{ij}$ variation}} & \multicolumn{2}{c}{\textbf{$r_{ij}$ variation}} \\ 
\cmidrule(lr){3-4} \cmidrule(lr){5-6} \cmidrule(lr){7-8}
 &  & $\mathcal{U}(0.1,0.3)$ & $\mathcal{U}(0.7,0.9)$ & $\mathcal{U}(0,0.1)$ & $\mathcal{U}(1,10)$ & $\mathcal{U}(0,0.1)$ & $\mathcal{U}(1,10)$ \\ 
\midrule
Random   & $0.17 \pm 1.73 \times 10^{-1}$ & $4.00 \pm 2.04 \times 10^{1}$ & $0.11 \pm 7.75 \times 10^{-3}$ & $0.10 \pm 7.75 \times 10^{-3}$ & $6.03 \pm 5.13$ & $0.15 \pm 2.24 \times 10^{-1}$ & $0.50 \pm 3.74 \times 10^{-1}$ \\
PPO      & $0.13 \pm 2.45 \times 10^{-2}$ & $2.10 \pm 7.36 \times 10^{0}$ & $0.07 \pm 4.47 \times 10^{-3}$ & $0.08 \pm 1.00 \times 10^{-2}$ & $3.68 \pm 1.22$ & $0.12 \pm 1.41 \times 10^{-2}$ & $0.25 \pm 8.37 \times 10^{-2}$ \\
DQN      & $0.15 \pm 7.75 \times 10^{-2}$ & $13.1 \pm 1.64 \times 10^{2}$ & $0.10 \pm 1.73 \times 10^{-2}$ & $0.12 \pm 2.83 \times 10^{-2}$ & $5.92 \pm 3.26$ & $0.15 \pm 7.07 \times 10^{-2}$ & $0.25 \pm 2.24 \times 10^{-1}$ \\
A2C      & $0.14 \pm 6.32 \times 10^{-2}$ & $2.22 \pm 9.82 \times 10^{0}$ & $0.11 \pm 2.17 \times 10^{-2}$ & $0.07 \pm 1.79 \times 10^{-2}$ & $3.62 \pm 2.93$ & $0.14 \pm 3.87 \times 10^{-2}$ & $0.43 \pm 2.74 \times 10^{-1}$ \\
\bottomrule
\end{tabular}%
}
\vspace{-18pt}
\end{table*}

\begin{figure}[tb]
    \centering
    \includegraphics[width=1.0\linewidth]{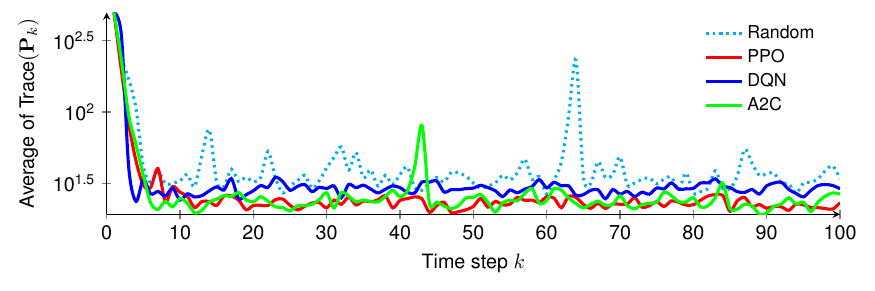}
    \vspace{-22pt}
    \caption{Comparison of the average estimation error at each step 
    over 100 steps (averaged across 1000 trials) 
    for different scheduling policies. }
    \label{fig:avg_P}
    \vspace{-10pt}
\end{figure}

\begin{figure}[tb]
    \centering
    \includegraphics[width=1.0\linewidth]{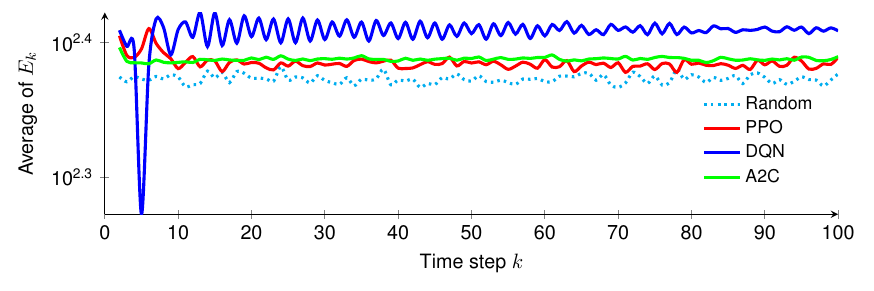}
    \vspace{-22pt}
    \caption{Comparison of the energy consumption
    at each step over 100 steps (averaged across 1000 trials) 
    for different scheduling policies.}
    \vspace{-13pt}\label{fig:avg_E}
\end{figure}

Figure~\ref{fig:avg_P} and Fig.~\ref{fig:avg_E} present the progression of the trace of the error covariance $\mathrm{trace}(\mathbf{P}_k)$ and the energy consumption $E_k$, respectively, over $100$ time steps. Random scheduling is inadequate in leveraging informative sensors optimally, leading to consistently elevated uncertainty despite minimal energy consumption over time. Conversely, PPO consistently reduces the estimation error more effectively than all other baselines while maintaining competitive energy usage. Its energy consumption surpasses that of random scheduling, which is logical since it signifies the necessity of utilizing more energy than the average to achieve the desired accuracy. This observation aligns with the theoretical trade-off established in Sec.~IV-B, where PPO efficiently balances the information gain $\Delta_i(\mathbf{P},\delta)$ against the costs associated with sensor energy. Meanwhile, DQN exhibits pronounced fluctuations, whereas A2C produces comparatively smoother curves; however, it does not attain the low-error regime achieved by PPO.


\subsubsection{Insights into Scheduling Dynamics}
\begin{figure}[tb]
\centering
\subfloat[Estimation uncertainty and energy usage trade off. For example, sensors 5 and 17 (circled in red) are selected during high-uncertainty phase, where PPO prioritizes highly informative and energy-efficient sensors to reduce estimation error rapidly.]{
    \includegraphics[width=0.97\linewidth]{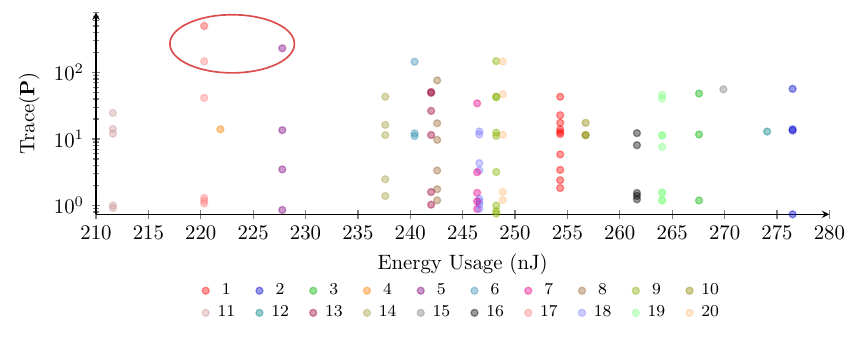}
    \label{fig:insight-a}
    }\\
    \vspace{-10pt}
\subfloat[Comparison of average delay $\bar{\delta}^i$ between theory and simulation]{
    \includegraphics[width=0.8\linewidth]{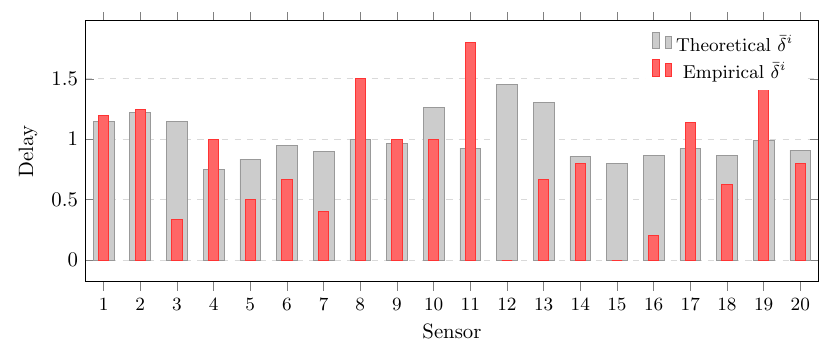}
    \label{fig:insights-b}
}
\caption{Demonstration in a single run of PPO-based policy in terms of optimizing and trade off between trace of error covariance, energy consumption, sensor heterogeneity, and information freshness.}
\label{fig:insights}
\vspace{-15pt}
\end{figure}

To better understand why PPO outperforms alternatives, we inspect a single representative run of the agent in Fig.~\ref{fig:insight-a}, revealing how PPO balances estimation uncertainty and sensor energy usage. 
\begin{itemize}[leftmargin=*]
    \item When $\mathrm{trace}(\mathbf{P})$ is large, PPO prioritizes highly informative sensors with relatively low transmission energy (e.g., sensor 5, 17), quickly reducing estimation error at minimal cost.
    \item As uncertainty decreases, PPO begins to schedule a broader set of sensors, suggesting it improves estimation across all five state dimensions, not only relying on the most informative sensors. At this
    stage, delay becomes more critical, the agent often trades off between informativeness, energy, and packet staleness, as represented by $\zeta_i(\mathbf{P}, \delta) = \Delta_i(\mathbf{P}, \delta)/\beta E^i$.
    \item Once estimation stabilizes, PPO reduces the number of active sensors and primarily schedules a small set of low-cost, moderately informative sensors. This shows that PPO learns to maintain accuracy with minimal energy expenditure.
\end{itemize}


Fig.~\ref{fig:insights-b} compares theoretical and empirical average delays. In most cases, PPO lowers the delay compared to theory. However, for a few sensors (e.g., sensor 8, 11, 17 and 19), PPO occasionally chooses an action that results in a higher delay. This highlights that PPO is not entirely conservative: it sometimes takes \say{risky actions} if the expected information gain outweighs the delay penalty. This observation confirms the theoretical expectation of an exploration, exploitation trade-off, while also showing that PPO remains primarily risk-averse overall.

\subsubsection{Robustness to Parameter Variations}
\label{sec:paramterVariations}
Finally, we examine the robustness of each scheduling policy to changes in system parameters. In Table~III, we vary (i) probability of new measurements, (ii)  process noise covariance, and (iii) measurement noise covariance. 

\paragraph{Measurement generation probability}
When $p_i$ decreases (measurements become less frequent and
delays increase), all methods perform worse, with higher mean
and variance of the objective. PPO and A2C adapt relatively
well, whereas DQN suffers significantly for its value-based nature being less suited to stochastic environments. Random
scheduling also degrades, but not as sharply as DQN.
When $p_i$ increases toward
$0.7$--$0.9$, results improve across all methods, approaching
the \textit{Standard} case.

\paragraph{Process noise}
Reducing $\mathbf{Q}$ makes the system model more accurate, which
stabilizes estimation. As expected, all methods show improved
performance with lower variance. Increasing $\mathbf{Q}$ makes the
system rely more heavily on sensor updates. Here, PPO and A2C still outperform, while DQN follows the same
relative order as in the \textit{Standard} case but with amplified
variance. This is consistent with theory, as actor-critic methods (e.g., PPO, A2C) are better at adapting to noise-dominated regimes.

\paragraph{Measurement noise}
When $\mathbf{R_i}$ decreases, measurements are more accurate,
making scheduling easier for all methods; conversely,
larger $\mathbf{R_i}$, reduces informativeness and degrades performance. Unlike $p_i$ and $\mathbf{Q}$, $\mathbf{R_i}$ variations affect
measurement quality directly rather than delay stochasticity.
Accordingly, PPO and A2C retain superiority but with smaller
relative gains, while DQN/random degrade as
expected.



\vspace{-3pt}

\section{Summary}
\label{section-VI}

This work introduced two main contributions: a delay-aware estimator that maintains stability under delayed updates, and a PPO-based scheduler that optimally balances accuracy, energy, heterogeneity, and freshness. Together, they form a unified framework for reliable remote state estimation in delay-prone wireless networks. Our analysis establishes conditions for possible stability, providing theoretical insight, while simulations demonstrate strong empirical performance and robustness compared to other baselines. Future work will extend this framework toward adaptive multi-objective optimization for diverse IoT and industrial applications.



\vspace{-5pt}
\bibliographystyle{IEEEtran}
\bibliography{ref/references}

@IEEEtranBSTCTL{IEEEexample:BSTcontrol,
  CTLuse_forced_etal       = "yes",
  CTLmax_names_forced_etal = "2",
  CTLnames_show_etal       = "1" 
}

@inproceedings{optuna_2019,
    title={Optuna: A Next-generation Hyperparameter Optimization Framework},
    author={Akiba, Takuya and Sano, Shotaro and Yanase, Toshihiko and Ohta, Takeru and Koyama, Masanori},
    booktitle={Proceedings of the 25th {ACM} {SIGKDD} International Conference on Knowledge Discovery and Data Mining},
    year={2019}
}

@misc{ao-insufficient,
      title={On the Role of Age and Semantics of Information in Remote Estimation of {M}arkov Sources}, 
      author={Jiping Luo and Nikolaos Pappas},
      year={2025},
      eprint={2507.18514},
      archivePrefix={arXiv},
      primaryClass={cs.IT},
      note={\url{https://arxiv.org/abs/2507.18514}}, 
}

@ARTICLE{application,
  author={Trigka, Maria and Dritsas, Elias},
  journal={IEEE Access}, 
  title={Wireless Sensor Networks: From Fundamentals and Applications to Innovations and Future Trends}, 
  year={2025},
  volume={13},
  number={},
  pages={96365-96399},
  doi={10.1109/ACCESS.2025.3572328}}

@book{dansimon,
author = {Simon, Dan},
title = {Optimal State Estimation: Kalman, H Infinity, and Nonlinear Approaches},
year = {2006},
isbn = {0471708585},
publisher = {Wiley-Interscience},
address = {USA}
}

@misc{ppo,
      title={Proximal Policy Optimization Algorithms}, 
      author={John Schulman and Filip Wolski and Prafulla Dhariwal and Alec Radford and Oleg Klimov},
      year={2017},
      eprint={1707.06347},
      archivePrefix={arXiv},
      primaryClass={cs.LG},
      url={https://arxiv.org/abs/1707.06347}, 
}

@INPROCEEDINGS{LTI-2,
  author={Zhou, Binquan and Wang, Zhuo and Zhai, Yueyang and Yuan, Heng},
  booktitle={IEEE DDCLS}, 
  title={Data-Driven Analysis Methods for Controllability and Observability of A Class of Discrete {LTI} Systems with Delays}, 
  year={2018},
  volume={},
  number={},
  pages={380-384},
  doi={10.1109/DDCLS.2018.8515909}}

@ARTICLE{aoi1,
  author={Chang, Taige and Cao, Xianghui and Zheng, Wei Xing},
  journal={IEEE Trans. Automat. Contr.}, 
  title={A Lightweight Sensor Scheduler Based on {AoI} Function for Remote State Estimation Over Lossy Wireless Channels}, 
  year={2024},
  volume={69},
  number={3},
  pages={1697-1704},
  doi={10.1109/TAC.2023.3328244}}

@INPROCEEDINGS{sensor-schedule-delay-2,
  author={Chandrasekaran, Sanjay and Varadan, Vishnu and Krishnan, Siva Vignesh and Dörfler, Florian and Mamduhi, Mohammad H.},
  booktitle={European Control Conference (ECC)}, 
  title={Distributed State Estimation for Linear Time-Varying Systems with Sensor Network Delays}, 
  year={2023},
  volume={},
  number={},
  pages={1-6},
  doi={10.23919/ECC57647.2023.10178246}}

@ARTICLE{energy-alone,
  author={Cao, Xianghui and Wang, Jia and Cheng, Yu and Jin, Jiong},
  journal={IEEE Internet Things J.}, 
  title={Optimal Sleep Scheduling for Energy-Efficient {AoI} Optimization in Industrial Internet of Things}, 
  year={2023},
  volume={10},
  number={11},
  pages={9662-9674},
  doi={10.1109/JIOT.2023.3234582}}

@ARTICLE{energy1,
  author={Yang, Wei and Durisi, Giuseppe and Polyanskiy, Yury},
  journal={IEEE Trans. Inf. Theory}, 
  title={Minimum Energy to Send $k$ Bits Over Multiple-Antenna Fading Channels}, 
  year={2016},
  volume={62},
  number={12},
  pages={6831-6853},
  doi={10.1109/TIT.2016.2615629}}

@article{Sun2025AoISchedulingTAC,
  author  = {Bowen Sun and Gy{\"o}rgy D{\'a}n and James Gross and Xianghui Cao},
  title   = {{AoI}-based Optimal Transmission Scheduling for Multi-process Remote Estimation},
  journal = {IEEE Trans. Automat. Contr.},
  year    = {2025},
  doi     = {10.1109/TAC.2025.3568094},
}


\end{document}